\documentclass[
onecolumn,
showkeys,
longbibliography,
nofootinbib,
superscriptaddress,
aps,
notitlepage
]{revtex4-1}
\usepackage[utf8]{inputenc}

\usepackage{amsmath}
\usepackage{amssymb}
\usepackage{pifont}
\usepackage{bm}
\usepackage{enumitem}
\usepackage[acronym]{glossaries}
\usepackage{graphicx}
\usepackage[colorlinks,unicode]{hyperref}
\usepackage[capitalise]{cleveref}  
\usepackage{mathtools}
\usepackage{multirow}
\usepackage{physics}
\usepackage{siunitx}
\usepackage{ulem}
\usepackage[dvipsnames,table,xcdraw]{xcolor}

\makeglossaries
\newacronym{dm}{DM}{dark matter}
\newacronym{ce}{CE}{Cosmic Explorer}
\newacronym{et}{ET}{Einstein Telescope}
\newacronym{snr}{SNR}{signal-to-noise ratio}

\makeatletter
\renewcommand\onecolumngrid{%
\do@columngrid{one}{\@ne}%
\def\set@footnotewidth{\onecolumngrid}%
\def\footnoterule{\kern-6pt\hrule width 1.5in\kern6pt}%
}
\makeatother

\renewcommand{\arraystretch}{1.4}

\newcommand\myshade{80}
\colorlet{mylinkcolor}{ForestGreen}
\colorlet{mycitecolor}{Red}
\colorlet{myurlcolor}{violet}

\hypersetup{
  linkcolor  = mylinkcolor!\myshade!black,
  citecolor  = mycitecolor!\myshade!black,
  urlcolor   = myurlcolor!\myshade!black,
  colorlinks = true
}

\definecolor{Green2}{RGB}{44, 160, 44}

\newcommand{\lab}[1]{{\mathrm{#1}}}

\DeclareSIUnit\solarmass{\ensuremath{\mathrm{M}_\odot}}
\DeclareSIUnit\parsec{pc}
\DeclareSIUnit\gigaparsec{Gpc}
\DeclareSIUnit\year{yr}

\newcommand{\GRAPPA}{Gravitation Astroparticle Physics Amsterdam (GRAPPA),\\ Institute for Theoretical Physics Amsterdam and Delta Institute for Theoretical Physics,\\ University of Amsterdam, Science Park 904, 1098 XH Amsterdam, The Netherlands}

\newcommand{\IFIC}{Instituto de F\'isica Corpuscular, Universidad de Valencia and CSIC,\\Edificio Institutos de Investigac\'ion, Calle Catedr\'atico Jos\'e Beltr\'an 2, 46980 Paterna, Spain}

\newcommand{\IFCA}{Instituto de F\'isica de Cantabria (IFCA, UC-CSIC), Avenida de Los Castros s/n, 39005 Santander, Spain}

\newcommand{\NBI}{Niels Bohr International Academy, Niels Bohr Institute, Blegdamsvej 17, 2100 Copenhagen, Denmark}

\newcommand{\ciela}{Ciela -- Computation and Astrophysical Data Analysis Institute, Montréal, Quebec, Canada}
\newcommand{\UdeM}{Département de Physique, Université de Montréal, 1375 Avenue Thérèse-Lavoie-Roux, Montréal, QC H2V 0B3, Canada}
\newcommand{\Mila}{Mila -- Quebec AI Institute, 6666 St-Urbain, \#200, Montreal, QC, H2S 3H1}

\begin{document}

\title{Disks, spikes, and clouds: distinguishing environmental effects on BBH gravitational waveforms}

\author{Philippa S. Cole}
\email{p.s.cole@uva.nl}
\affiliation{\GRAPPA}

\author{Gianfranco Bertone}
\affiliation{\GRAPPA}

\author{Adam Coogan}
\affiliation{\ciela}
\affiliation{\UdeM}
\affiliation{\Mila}

\author{Daniele Gaggero}
\affiliation{\IFIC}

\author{Theophanes Karydas}
\affiliation{\GRAPPA}

\author{Bradley J. Kavanagh}
\affiliation{\IFCA}

\author{Thomas F. M. Spieksma}
\affiliation{\GRAPPA}
\affiliation{\NBI}

\author{Giovanni Maria Tomaselli}
\affiliation{\GRAPPA}

\begin{abstract}
Future gravitational wave interferometers such as LISA, Taiji, DECIGO, and TianQin, will enable precision studies of the environment surrounding black holes. 
In this paper, we study intermediate and extreme mass ratio binary black hole inspirals, and consider three possible environments surrounding the primary black hole: accretion disks, dark matter spikes, and clouds of ultra-light scalar fields, also known as gravitational atoms. We present a Bayesian analysis of the detectability and measurability of these three environments. Focusing for concreteness on the case of a detection with LISA, we show that the characteristic imprint they leave on the gravitational waveform would allow us to identify the environment that generated the signal, and to accurately reconstruct its model parameters.  
\end{abstract}


\maketitle

\section{Introduction}

The next generation of gravitational wave detectors are expected to come online in the 2030s. Those set to explore the milli- and deci-Hertz regimes, such as LISA~\cite{Baker:2019nia}, Taiji~\cite{2021PTEP.2021eA108L}, DECIGO~\cite{Kawamura:2020pcg} and TianQin~\cite{TianQin:2015yph}, will open a new window for gravitational wave discoveries. They will have a much lower frequency range than the current LIGO~\cite{2015CQGra..32g4001L}, Virgo~\cite{VIRGO:2014yos} and KAGRA~\cite{KAGRA:2020tym} detectors. For example, LISA is expected to be sensitive in the range $10^{-4}-1\,\mathrm{Hz}$, meaning that black hole (BH) binaries with much larger chirp masses will be detectable. Moreover, these sources will stay in band for long durations, up to weeks, months or years in some cases, especially for intermediate mass ratio inspirals (IMRIs) and extreme mass ratio inspirals (EMRIs), which take longer to inspiral than more equal-mass binaries. 
Observations of IMRIs and EMRIs provide a unique opportunity to learn about the environments of the binaries~\cite{Macedo:2013qea,Barausse:2014tra,Barausse:2014pra}. This is because not only will the binaries stay in the sensitive range of the detector for a considerable amount of time, allowing the imprints of environmental effects to accumulate in the gravitational waveform, but also the environment of the central BH is more robust to disruptions by a much lighter companion object~\cite{Berry19}.

In this paper, we compare environmental effects on intermediate mass ratio binaries in the milli-Hertz band in three different scenarios, namely accretion disks~\cite{tanaka2002,Derdzinski:2018qzv,Duffell:2019uuk,Derdzinski_2020,antonelli22}, cold dark matter (CDM)  spikes~\cite{Gondolo:1999ef,Bertone:2005xz,Eda_2013,Eda_2015,Yue:2018vtk,Kavanagh:2020cfn,Coogan:2021uqv}, and clouds of ultra-light scalar fields~\cite{Dolan:2007mj,Arvanitaki:2009fg,Arvanitaki:2010sy,Brito:2015oca,Baumann:2019eav,Baumann:2019ztm,Baumann:2021fkf,Baumann:2022pkl}. 
We are predominantly interested in learning about the nature of dark matter (DM) from the gravitational waveform, and hence our focus is on the modelling of CDM density spikes around black holes and clouds of ultra-light scalar fields produced by superradiance -- a system which is otherwise known as a gravitational atom. However, accretion disks can act on the waveform in the same way as these dark spikes and clouds, so it is vital to determine whether there is a chance of confusion between DM and baryonic effects.

\section{Modelling the environments}\label{sec:model-env}

\begin{figure}[t]
    \centering
    \includegraphics[width=0.48\textwidth]{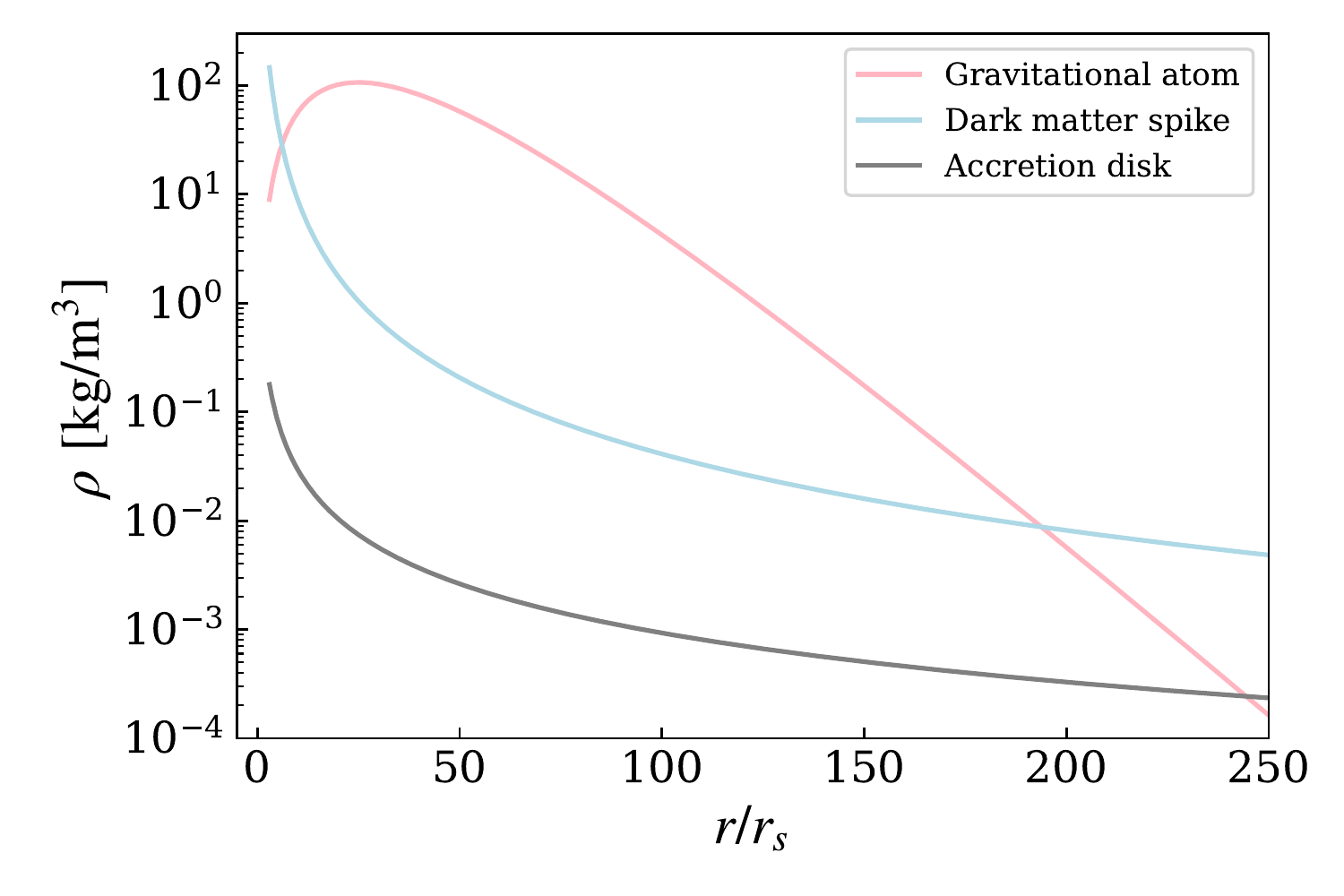}
    \includegraphics[width=0.48\textwidth]{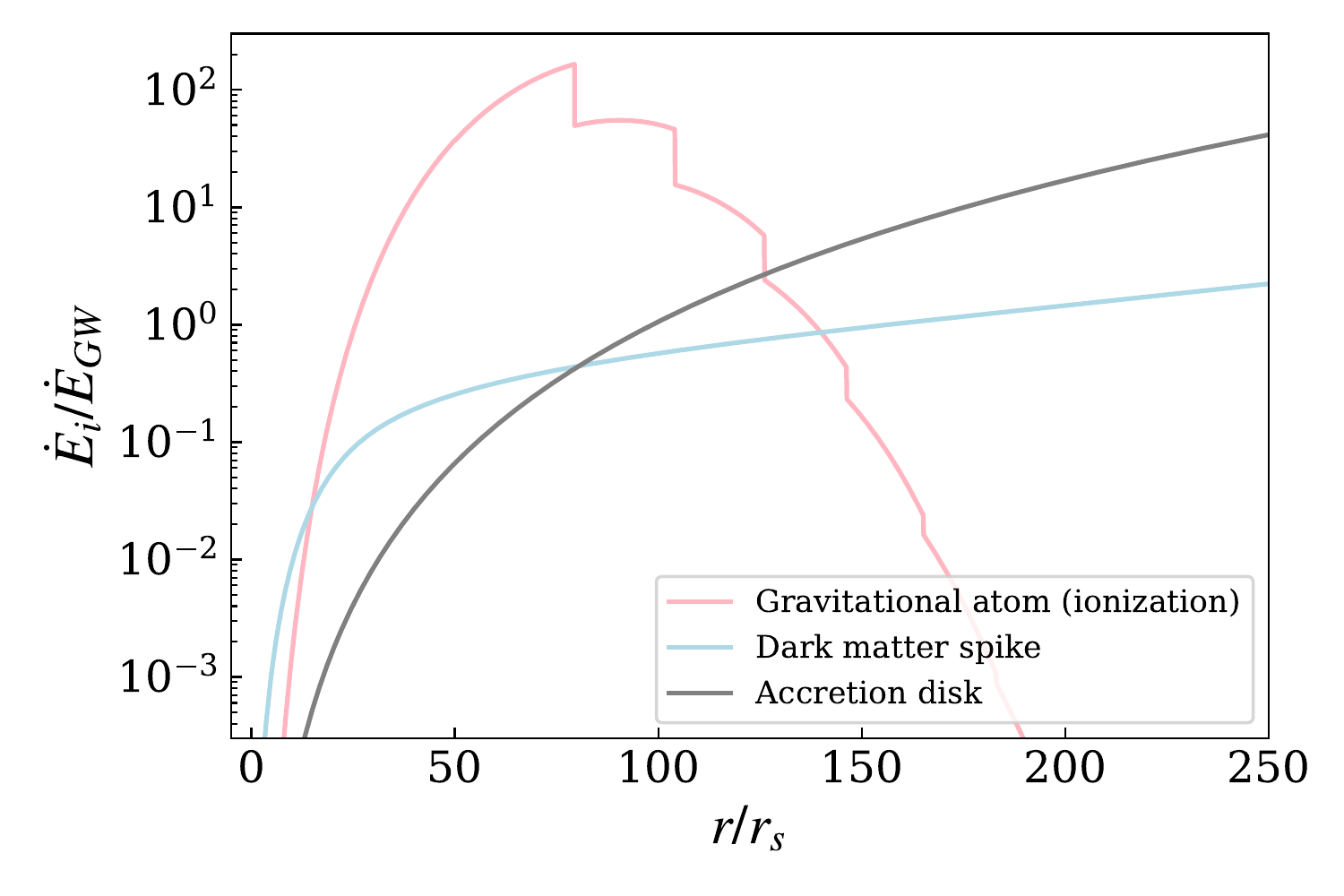}
    \caption{Left: Initial density profiles of environments around a $10^5\,{\rm M_\odot}$ black hole. Right: Energy losses due to environment normalised by the energy losses due to gravitational waves.}
    \label{fig:profile}
\end{figure}

We study three possible environments for intermediate and extreme mass ratio binary BH inspirals that may have an observable effect on the gravitational waveform: accretion disks, dark matter spikes and clouds of ultra-light scalar fields, also known as gravitational atoms. These environments can be characterised by their density profiles around the central BH with mass $m_1$, as shown in the left panel of figure \ref{fig:profile}.

\noindent
\paragraph{Cold collisionless dark matter.}
We model the initial density profile of the CDM spike with a power law:
\begin{equation}
\label{eq:DMspike_profile}
    \rho_\mathrm{CDM}(r)=\rho_6\left(\frac{r_6}{r}\right)^{\gamma_s}\,,
\end{equation}
where $\rho_6$ is the density of the spike at a reference distance of $r_6=10^{-6}\,{\rm pc}$ from the central BH and $\gamma_s$ is the slope of the spike. For spike formation from the adiabatic growth of an IMBH at the centre of a DM halo with an initial slope of $\gamma_i$, the final slope of the spike will be $\gamma_s = (9-2 \gamma_i)/(4-\gamma_i)$~\cite{Gondolo:1999ef}. For typical values of $\gamma_i \in[0, 2]$, this gives $\gamma_s \in [2.25, 2.5]$, with a value of $\gamma_s = 7/3$ for an initial NFW profile, which we assume here.

\noindent
\paragraph{Gravitational atom.}

The ultralight boson cloud surrounding the central BH is assumed to be in a pure $|n\ell m\rangle$ eigenstate, with wavefunction
$
    \psi(t,\vec r)=R_{n\ell}(r)Y_{\ell m}(\theta,\phi)e^{-i(\omega_{n\ell m}-\mu) t},
$
where $Y_{\ell m}$ are spherical harmonics and $R_{n \ell}(r)$ are the hydrogenic radial functions as laid out explicitly in \cite{Baumann:2021fkf}. Additionally, $\psi$ is related to the scalar field $\Phi$ via $\Phi=\psi e^{-i\mu t}/\sqrt{2\mu}$, where $\mu$ is the mass of the scalar field. This model is valid under the assumption of $\alpha/\ell\ll1$, where $\alpha\equiv Gm_1\mu$ is the so-called gravitational fine structure constant, in which case the cloud is mostly non-relativistic. The mass density can then be defined as
\begin{equation}
    \rho(\vec r)=M_\lab{c}|\psi(\vec r)|^2,
\end{equation}
where $M_\lab{c}$ is the total mass of the cloud. If $\Phi$ is a real field, rather than a complex one, $\psi$ has to be replaced with $2\Re[\psi]$ in its relation to $\Phi$ and $\rho$. The value of $M_\lab{c}$ is determined by the mass and spin of the black hole before the superradiant instability formed the cloud, and can reach a maximum of about $10\%$ of the central BH mass. We will consider $M_\lab{c}$ as an independent parameter because other processes, like the decay of the cloud into GWs \cite{Yoshino:2013ofa}, can change its value.

\noindent
\paragraph{Accretion disk.}
We model a locally isothermal disk, which is equivalent to a locally constant speed of sound and therefore a locally constant Mach number $M=r/h$, where $h$ is the scale height of the disk. Given that we are interested in very dense environments in order for environmental effects to cause a significant dephasing, we will focus on thin disks such that ${M} \gg 1$.
In terms of the surface density of the disk, we use the same parametrisation as in \cite{Derdzinski_2020}, so as to be in the regime where analytical expressions for gas torques in accretion disks have been calibrated with numerical simulations, as will be discussed in \cref{sec:elosses}. The surface density is described as a static power-law profile
\begin{equation}
    \Sigma(r)=\Sigma_0\left(\frac{r}{3r_s}\right)^{-1/2}\,,
\end{equation}
where $\Sigma_0$ is the surface density normalisation, $r_s$ is the Schwarzschild radius of the central black hole and the slope has been fixed. See \cite{antonelli22} for the case of a varying slope. Finally, we estimate the volume density of the disk for the purpose of comparison with the other environments in the left hand panel of \cref{fig:profile} with $\rho(r)=\Sigma(r)/2h$.

\section{Energy losses and dephasing of the waveform}
\label{sec:elosses}

The evolution of the binary's inspiral depends on the rate of energy loss of the system. We assume that the companion ${\color{violet}m_2}$ moves slowly inwards on quasi-circular orbits and that energy balance is satisfied with $\dot{E}_{\rm orb} = -\dot{E}_{\rm GW}-\dot{E}_{\rm env}$, where we use the Keplerian expression for the orbital energy $E_{\rm orb}=-{Gm_1m_2}/{(2r)}$ and model the GW energy losses at Newtonian order. We model the energy losses induced by the environment $\dot{E}_\mathrm{env}$ as a linear combination of the relevant effects for each system, including dynamical friction (DF), ionization, torques, and accretion onto the companion object, all of which are described in detail in \cref{app:dd,app:ga,app:torques} and briefly summarized in this section.

The relative importance of the environmental effects with respect to the energy radiated away due to gravitational waves is shown in the right panel of \cref{fig:profile}, where we plot $\dot{E}_{\rm env}/\dot{E}_{\rm GW}$ as a function of the separation of the binary in units of Schwarzschild radii for the benchmark parameters given in \cref{sec:benchmarks}. In each case, we work under the assumption of a small mass ratio $q=m_2/m_1<10^{-2.5}$ such that we do not expect the environment to be destroyed within the first few close encounters of the binary, and also so that the companion object can be treated as a point mass without an environment of its own.

\noindent
\paragraph{Cold collisionless dark matter.}

We model the effects of a DM spike, including feedback on the spike itself, following Refs.~\cite{Kavanagh:2020cfn,Coogan:2021uqv}. In this framework, the effect of accretion is sub-dominant, so the energy losses are solely due to dynamical friction, $\dot{E}_\mathrm{env}=\dot{E}_\mathrm{DF}$, which takes the form~\cite[Appendix~L]{BinneyAndTremaine}:
\begin{equation}
    \dot{E}_\mathrm{DF} = \frac{4\pi G^2 m_2^2 \rho_\mathrm{CDM}(r, t) \xi(v) \log\Lambda}{v}\,,
\end{equation}
where $v$ is the orbital velocity, $\xi(v)$ is the fraction of DM particles moving more slowly than $v$ and $\log\Lambda$ is the Coulomb logarithm, which encodes information about the minimum and maximum impact parameters relevant for the dynamical friction force.

\noindent
\paragraph{Gravitational atom.}
The case of the gravitational atom is modelled similarly to \cite{Baumann:2021fkf}. 
The orbits are assumed to be 
lying on the equatorial plane defined by the spin of the central black hole and of the cloud, and we choose for the companion to be co-rotating with the cloud. 
Besides GW emissions, two effects are taken into account for the binary evolution: the ``ionization'' of the cloud due to the gravitational perturbation of the secondary, and the accretion of the cloud by the smaller black hole moving through it\footnote{The energy is not conserved because accretion is a dissipative process. The balance of angular momentum, however, can be written in a similar form, where accretion contributes as an additional ``force'', see (\ref{eqn:drdt-atom}).} so we have $\dot{E}_\mathrm{env}=\dot{E}_\mathrm{ion} + \dot{E}_\mathrm{acc}$.

\noindent
\paragraph{Accretion disk.}\label{sec:elosses_AD}

In the case of accretion discs, the dominant cause of dephasing for compact binaries arises from gas torques (see \cref{app:torques}).
In analogy with so-called Type-I planet migration, we write the total net torque \textit{on the secondary black hole}, with mass much smaller than the primary black hole, coplanar with, and fully embedded in, an accretion disk, as~\cite{Derdzinski_2020}:
\begin{equation}\label{eq:torque}
    T_0=-\Sigma(r)r^4\Omega^2q^2{M}^2,
\end{equation}
where ${M}$ is the Mach number of the disk, $\Sigma(r)$ is the surface density of the (unperturbed) disk, $\Omega$ is the orbital angular velocity, and $q$ is the mass ratio (see the discussion in Appendix~\ref{app:torques}). Note that the negative sign makes torques act in the same direction as dynamical friction, so they lead to a faster inspiral with respect to the vacuum case.
We can write the energy losses due to gas torques $T_0$ in a differentially rotating accretion disk as
\begin{equation}
    \dot{E}_\mathrm{torque}=\frac{G^\frac{1}{2}T_0m_1}{4r^\frac{3}{2}(m_1+m_2)^\frac{1}{2}}.
\end{equation}

\section{Parameter inference to distinguish between environments}\label{sec:PE}

\subsection{Benchmark system}
\label{sec:benchmarks}

We study a black hole binary system with masses  $m_1=10^5\,\mathrm{M_\odot}$, $m_2=\SI{10}{\solarmass}$ and hence chirp mass $\mathcal{M}_{c,0}\simeq398\,\mathrm{M_\odot}$. We choose this $m_1$ because there are plausible formation scenarios for all three environments around a central object of this mass -- see e.g.~\cite{Gondolo:1999ef} for cold, collisionless dark matter, \cite{Greene_2020} for accretion disks, and \cite{Arvanitaki:2010sy,Hannuksela_2019} for gravitational atoms. Furthermore, the vacuum ISCO frequency of this system, $f_\mathrm{ISCO} = 0.044\,\mathrm{Hz}$ lies close to the bucket of the LISA noise curve, meaning that inspirals will take place in a frequency range where the detector has high sensitivity. Lastly, we choose a small mass ratio, $q=m_2/m_1=10^{-4}$, such that we do not expect the environments to be disrupted significantly by the companion object, and hence various assumptions which rely on this when calculating the energy losses for each environment hold.

The benchmark parameters we choose for each environment are as follows. For the dark dress, $\rho_6=\SI{1.17e17}{M_\odot/pc^3}$ and $\gamma_s=7/3$ \cite{Gondolo:1999ef}, for the accretion disk, $\Sigma_0{M}^2 = \SI{1.5e10}{kg/m^2}$, and for the gravitational atom, $\alpha=0.2$ and $M_\mathrm{c}/m_1=0.01$ \cite{Baumann:2021fkf,Baumann:2022pkl}. The value of $\Sigma_0{M}^2$ we choose for the accretion disk is unrealistically high (see e.g. \cite{Jiang_2019} for recent observations), but since we are mainly concerned with confusing a dark matter spike for an accretion disk, we want to show that when the effect of the dephasing is comparable, we can still distinguish between the environments. The signal-to-noise ratio (SNR) loss with respect to the best-fit vacuum signal is non-negligible (see \cref{fig:SNRloss}), providing a conservative comparison with the dark dress and gravitational atom. Lower and more realistic values of $\Sigma_0{M}^2$ would be more easily differentiable. Note that increasing any one of these parameters at a time increases the amount of dephasing with respect to the vacuum system (see \cref{fig:SNRloss} which will be discussed in \cref{sec:PE}).

\subsection{Parameter estimation with correct model}

Firstly, we demonstrate that we can reconstruct the parameters of each environment from the gravitational waveform of a detected 1-year duration signal, if matched filtering using a template bank with the correct parameters is used. We use the final year of the signal pre-merger, and fix the luminosity distance at $d_L=3.3 \,\mathrm{Gpc}$ such that the SNR is 15 for each system. We run parameter estimation using the nested sampling~\cite{nsskilling2004,nsskilling2006,multinest} code \texttt{dynesty}~\cite{dynesty}, with the log-likelihood given by the match integral between the sky- and polarization-angle averaged signal $d(t)$ and template $h(t)$, maximized over the extrinsic parameters (see \cref{sec:match}).

\renewcommand{\arraystretch}{1.3}
\renewcommand{\tabcolsep}{8pt}
\begin{table}[]
\begin{tabular}{lllll}
\toprule
                   & $\mathcal{M}_c$ & $\log_{10}(q)$ & $\theta_\mathrm{env}$                                                                                     &  \\ \cline{1-4}
Dark dress         & $\mathcal{M}_{c,0} \pm0.01\,\mathrm{M}_\odot$                & $\mathcal{U}(-5 , -2.5)$                & \begin{tabular}[c]{@{}l@{}}$\rho_6 = \mathcal{U}(0,10^{22})\,\mathrm{M_\odot/pc^3}$\\ $\gamma_s = \mathcal{U}(2,2.5)$\end{tabular} &  \\[0.5cm] 
Accretion disk     & $\mathcal{M}_{c,0}\pm0.005\,\mathrm{M}_\odot$                & N/A                                    & $\Sigma_0\mathcal{M}^2 = 10^{\mathcal{U}(8,16)}\,\mathrm{kg/m^2}$                                                                   &  \\[0.3cm] 
Gravitational atom & $\mathcal{M}_{c,0}\pm0.01\,\mathrm{M}_\odot$                & $\mathcal{U}(-5 , -2.5)$                & \begin{tabular}[c]{@{}l@{}}$\alpha=\mathcal{U}(0.1,0.4)$\\ $M_\mathrm{c}/m_1=10^{\mathcal{U}(-8,-2)}$\end{tabular}                         &  \\ 
\botrule
\end{tabular}
\caption{Prior ranges used for parameter estimation carried out using nested sampling. Those for the chirp mass $\mathcal{M}_c$ are narrow because the full prior volume drops out of the Bayes factor calculation when comparing models. All posteriors are contained within the priors except for $\gamma_s$.}
\label{tab:priors}
\end{table}

\begin{figure}[h!]
     \centering
     \includegraphics[width=\textwidth]{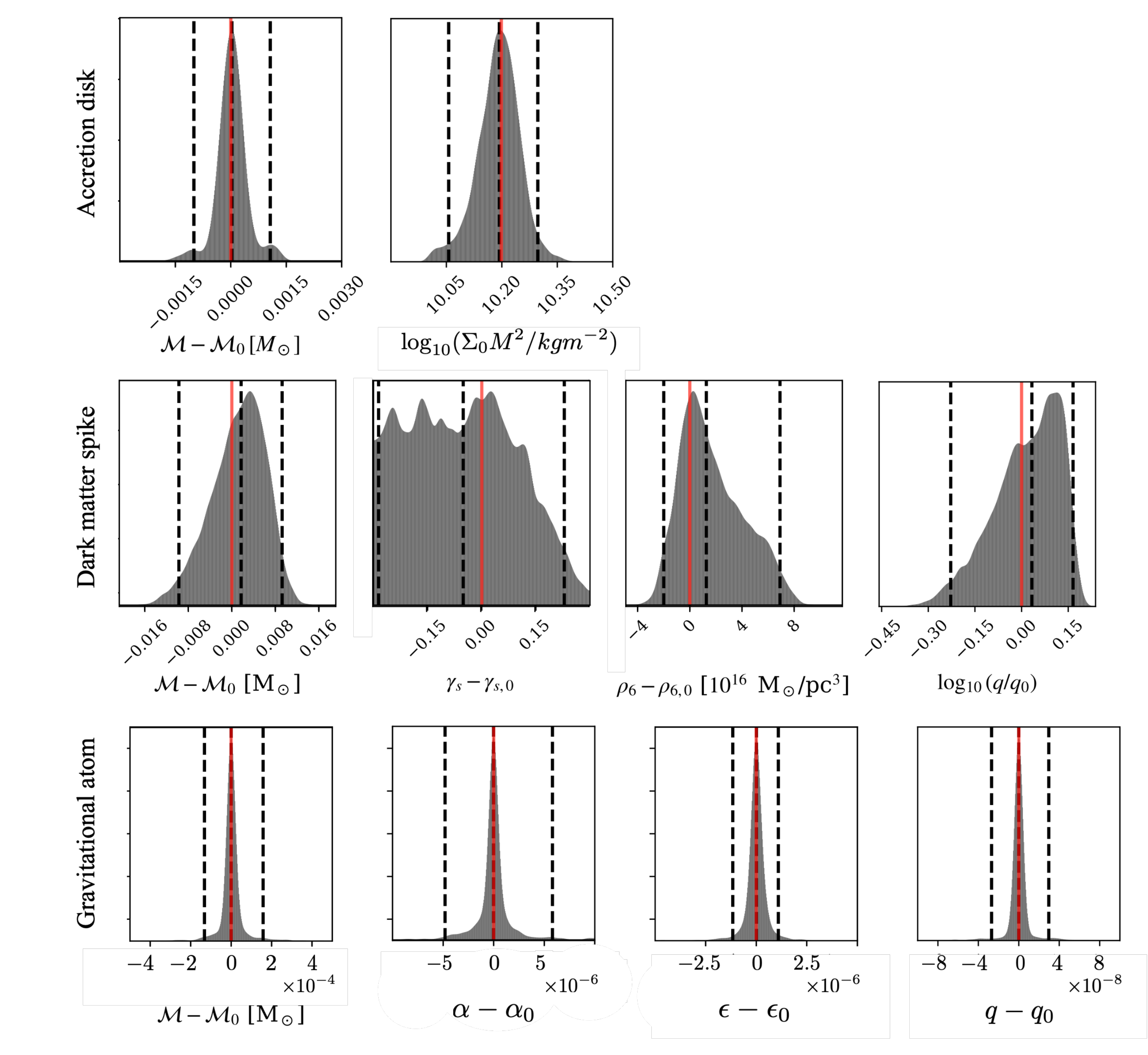}
        \caption{1-D posterior distributions for intrinsic and environmental parameters of an accretion disk (top row), dark dress (middle row) and gravitational atom (bottom row) signals with 1 year durations using the correct template in each case. Note that we use notation $\epsilon=M_c/m_1$ for brevity.}
\label{fig:correct}
\end{figure}

The posteriors for the intrinsic and environmental parameters are shown in \cref{fig:correct} for the dark dress, accretion disk and gravitational atom. All posteriors are smoothed with a 2\% Gaussian kernel. The red lines show the true values of the signal, whilst the vertical dashed lines show the 95\% (i.e., $2\sigma$) credible intervals. Intrinsic and environmental parameters are measured to excellent precision for all three environments with the exception of $\gamma_s$, for which longer-duration signals are required (see \cref{app:duration}). The precision of the measurements in the case of the gravitational atom is better than the systematic uncertainties in the waveform model. This shows that there is very little degeneracy between these parameters and prospects for measuring them from data are very hopeful, since orders of magnitude degradation in precision of the measurement would still lead to confident parameter inferences. Based on these extremely narrow posteriors for one year's worth of data, we also show the posterior distributions for just 1 month's worth of data in \cref{app:duration}, where all parameters are still very accurately measured with a degradation of the 95\% credible intervals by approximately an order of magnitude. Note that the mass ratio $q$ cannot be individually measured in the case of the accretion disk, because it appears in combination with $\Sigma_0 M^2$ in the dephasing contribution. This also explains the slightly better precision in the chirp mass measurement for the accretion disk over the dark dress, because we have fixed the mass ratio to its true value of $10^{-4}$.

Having shown that we can precisely measure the parameters of each system using the correct model in each case, we now go on to test whether it is possible to fit each system with an incorrect model.

\subsection{Distinguishing between environments}

\begin{figure}[h!]
     \centering
     \includegraphics[width=0.8\textwidth]{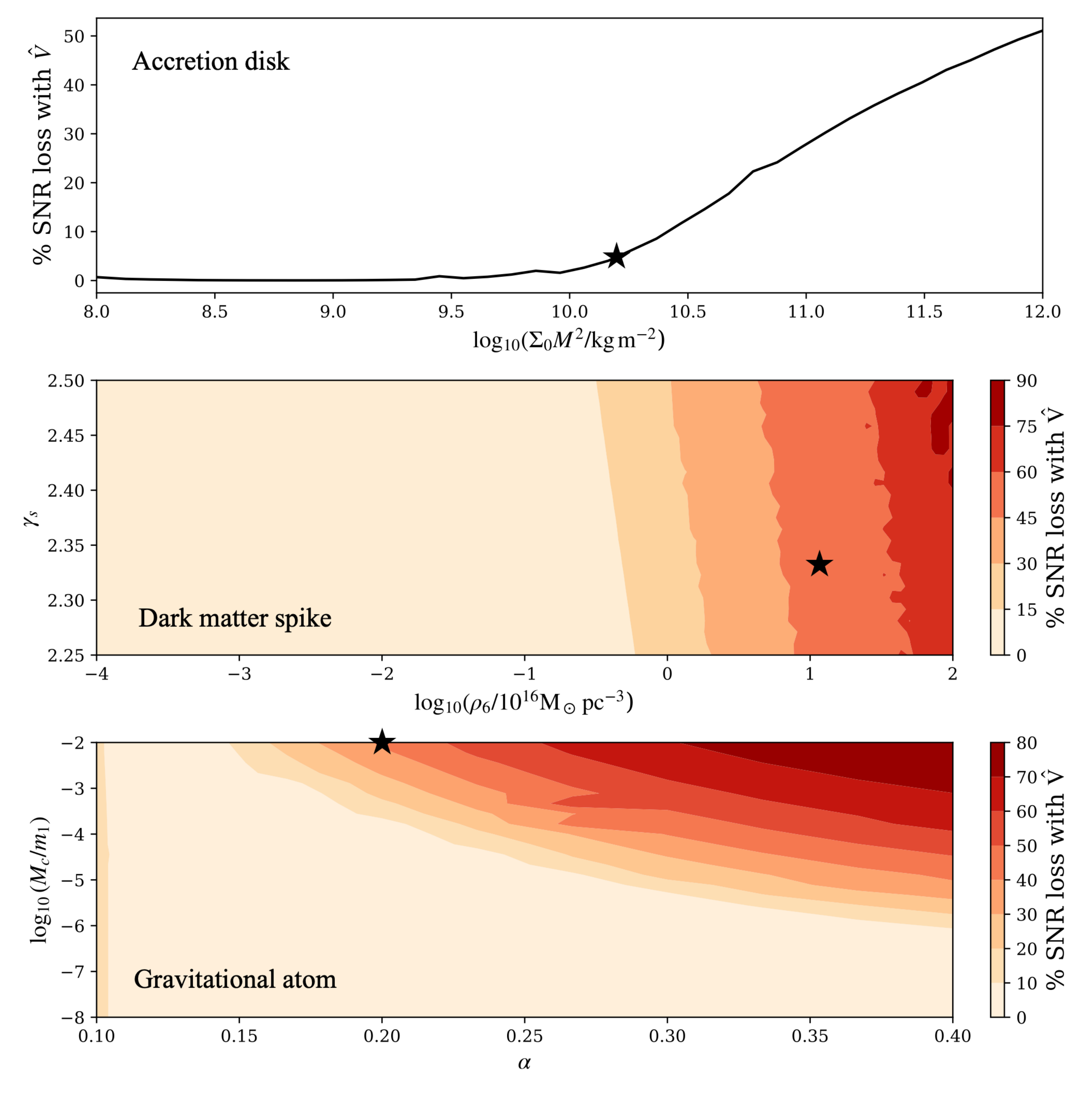}
     \caption{SNR loss with respect to best-fit vacuum system for accretion disk (top), dark dress (middle) and gravitational atom (bottom) signals with 1 year durations. Black stars indicate the benchmark parameters in each case.}
     \label{fig:SNRloss}
\end{figure}

Current gravitational wave template banks use only vacuum waveforms, so we first demonstrate that we can distinguish each environmental signal from the corresponding best-fit vacuum case. We diagnose the regions of the parameter space for each system where it might be possible to fit an environmental signal with a biased vacuum template by calculating the SNR lost between the signal and template waveforms. As a rule of thumb, we expect SNR losses of more than $30\%$ to compromise the ability to detect the signal with an incorrect template, and systems which incur small SNR losses we expect to lead instead to biased parameter estimation. The SNR loss results are shown in \cref{fig:SNRloss} for best-fit vacuum templates.

For the dark dress, a system with the well-motivated benchmark parameters which we measured in the previous section incurs SNR losses of order $50\%$. 
For the accretion disk, the SNR lost between the signal and the best-fit vacuum is $5\%$, which serves to provide a conservative comparison with the dark dress and the gravitational atom, since more realistic and lower values of $\Sigma_0 M^2$ with lower SNR loss would be more easily distinguishable. For the gravitational atom, a system with $\alpha=0.2$ and a conservative cloud mass of $1\%$ of the black hole mass leads to SNR loss of $\sim40\%$. Larger values of $\alpha$ lead to larger SNR loss, while decreasing the mass of the cloud relative to the black hole mass leads to smaller SNR loss as this approaches the vacuum regime.

\begin{figure}[t]
    \centering
    \includegraphics[width=0.325\textwidth]{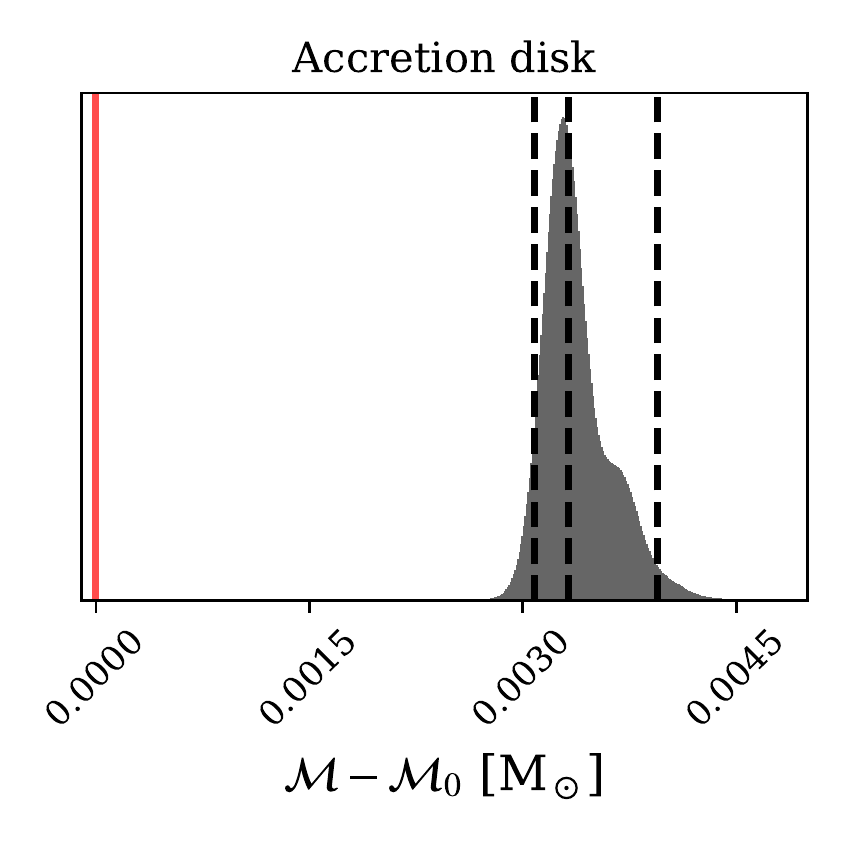}\hspace{0.25cm}
    \includegraphics[width=0.325\textwidth]{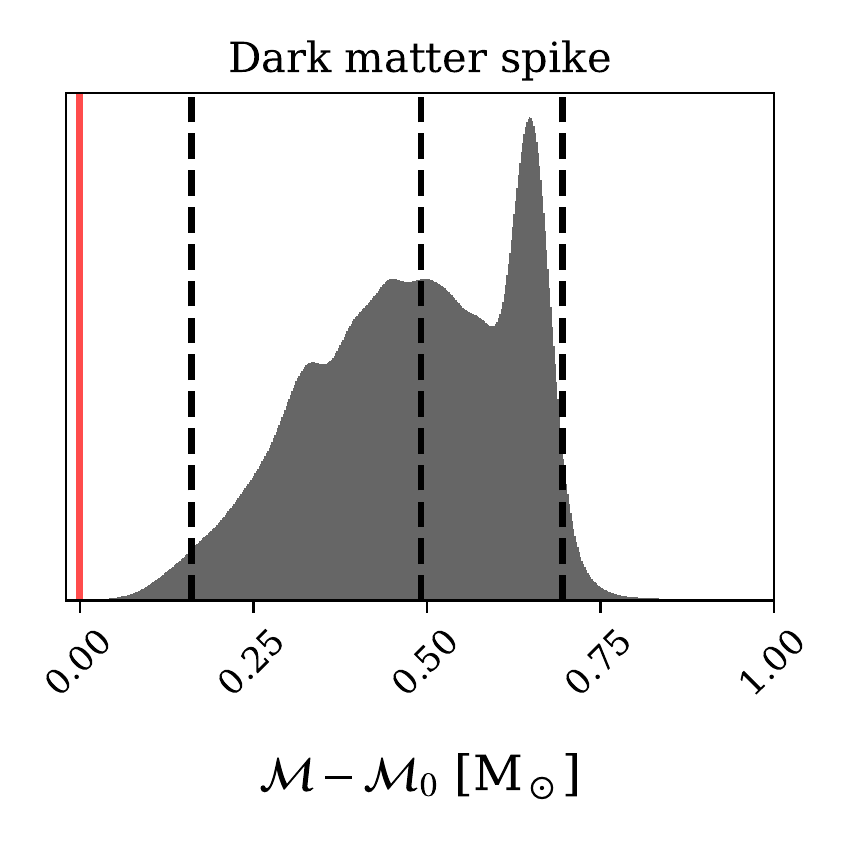}
    \includegraphics[width=0.325\textwidth]{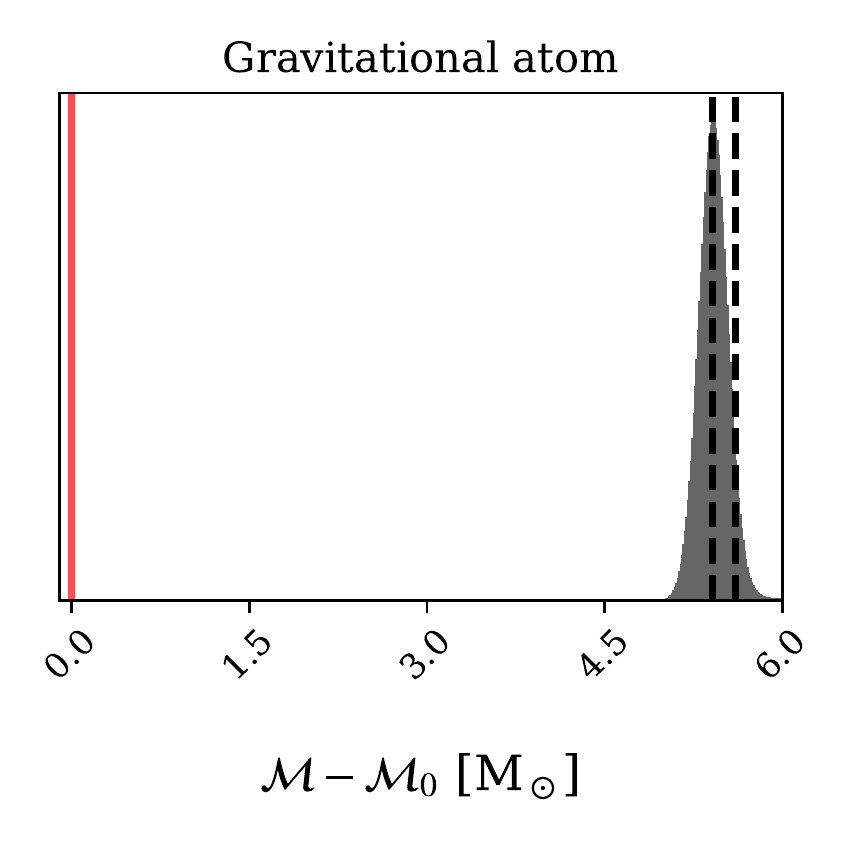}
    \caption{Posterior distribution for the chirp mass of an accretion disk signal (left), dark dress signal (middle) and gravitational atom signal (right) when fitted with a vacuum template, all with one year durations. The true chirp mass, $\mathcal{M}_0$, is indicated by the red vertical line at $\mathcal{M}-\mathcal{M}_0=0$.}
    \label{fig:vac_PE}
\end{figure}

To demonstrate that these SNR losses produce biased parameter inferences, and lack of Bayesian evidence for using the incorrect template for high SNR-loss systems, we run parameter estimation for the benchmark systems using nested sampling. The posteriors for the chirp mass, i.e.~the only free intrinsic parameter for a GR-in-vacuum waveform in our setup, are shown in figure \ref{fig:vac_PE}. When a vacuum template is used, the chirp mass for the accretion disk system is shifted from its true value by $3.3\times10^{-3}\,\mathrm{M_\odot}$, for the dark dress by $0.49\,\mathrm{M_\odot}$, and for the gravitational atom by $5.4\,\mathrm{M_\odot}$. This is explained by a larger chirp mass mimicking the speed-up of the inspiral due to the environmental effects. 

We calculate Bayes factors to compare the evidence for the correct models which include the environmental effects, versus vacuum. The Bayes factor is defined as the ratio of the evidence $p(d|D)$ for a signal $d$ under two different models (here $A$ and $B$):
\begin{equation}
    \mathcal{B}(d) = \frac{p(d|A)}{p(d|B)} \, .
\end{equation}
For a model $A$ with parameters $\theta$, the evidence is defined as
\begin{equation}
    p(d | A) = \int \mathrm{d} \theta\, p(d|h_\theta) \, p(\theta) \, ,
\end{equation}
where $p(\theta)$ is the prior on the model parameters, $h_\theta(t)$ is the waveform corresponding to parameters $\theta$ and $p(d | h_\theta)$ is the likelihood describing how probable the data is under that waveform model. We can extract an estimate for the evidence using nested sampling. We find $\log_{10}\mathcal{B}={34}$ for the dark dress, $\log_{10}\mathcal{B}={6}$ for the accretion disk, and $\log_{10}\mathcal{B}={39}$ for the gravitational atom, demonstrating undeniable support for the correct model in each case, even though there are clean posteriors for the chirp mass with a vacuum template in each case. Systems with parameter values that lead to higher SNR loss with respect to vacuum will lead to even larger Bayes factors, and it is unlikely that such systems will be detectable at all by matched filtering searches using only vacuum templates.

Finally, we ascertain whether we can distinguish between environments by computing the Bayes factors to compare each non-vacuum environment with every other one. We compute these Bayes factors for the same benchmark systems. The results are summarised in \cref{tab:BF}.

\begin{table}[b!]
    \begin{tabular}{|l|l|l|l|l}
    \cline{1-4}
             & Dark dress signal & Accretion disk signal & Gravitational atom signal &  \\ \cline{1-4}
    Vacuum template                                     & ${34}$                                 & ${6}$                                     & ${39}$                                         &  \\ \cline{1-4}
    Dark dress template         & -                                         & ${3}$                                     & ${39}$                                         &  \\ \cline{1-4}
    Accretion disk template     & ${17}$                                 & -                                             & ${33}$                                         &  \\ \cline{1-4}
    Gravitational atom template &                                       ${24}$    &       ${6}$                                        & -                                                 &  \\ \cline{1-4}
    \end{tabular}
    \caption{Logarithm of the Bayes factors, $\log_{10}\mathcal{B}$, comparing the evidence for the correct template that fits the signal, with an incorrect template. }
    \label{tab:BF}
\end{table}

The Bayes factors are very large, orders of magnitude larger than the $\mathcal{B}\sim\mathcal{O}(100)$ threshold for `confident' Bayesian preference for one model over another~\cite{jeffreys1998theory,10.2307/2291091}. This shows that we can confidently distinguish between environments when we compare the evidence for the parameter inference on a given signal with each environmental template bank. The driving force for this distinguishability is the way that the environments' waveforms evolve as a function of time, which makes it difficult to mimic the waveform of one environment by varying the parameters of another.

We estimate by how much these Bayes factors will degrade with systematic uncertainties on the template waveforms by re-calculating the evidence for the correct template with a multiplicative factor on the phase of the signal. In this case, the overall phase scaling of the dark dress system should be known to better than $0.1\%$ precision in order to confidently distinguish the system from other environments, while for the gravitational atom, the phase scaling should be known to better than $0.01\%$ precision. For the accretion disk, since we fix the mass ratio and therefore it is difficult for the other two parameter values to mimic a shift in the phase, the signal waveform needs to be completely contained within the template bank in order to achieve the Bayes factors in \cref{tab:BF}.

Furthermore, performing parameter inference including the extrinsic parameters, as well as using post-Newtonian waveforms including parameters like the effective spin that are not present in our analysis, will also likely decrease the Bayes factors. However, we emphasise the relative difference in the Bayes factors is extremely large, and that we can confidently distinguish between environments based on this.

\section{Conclusions}
\label{sec:conc}

Measuring the properties of the environments of intermediate mass ratio inspirals will be possible with next generation gravitational wave detectors. We have demonstrated that we can accurately reconstruct the parameters describing dark matter spikes, accretion disks, and gravitational atoms around an intermediate mass black hole, given a signal detected with an SNR of 15 of one year's duration. We have also shown that we can confidently distinguish between environments based on comparing the Bayesian evidence for using the correct environmental template for a given signal with an incorrect one. The correct environmental template is always strongly preferred, showing that we will not be at risk of misinterpreting an environmental signal for either a biased vacuum system, or the wrong type of environment. Furthermore, we show that SNR losses can be significant if the wrong template is used to fit the signal, and we therefore conclude that it is vital that environmental effects are taken into account when searching for and analysing long-duration signals from future gravitational wave detectors.

This work serves as a proof of concept for distinguishing between environments, and as a starting point for future refinements, including more realistic data analysis strategies as well as complete waveforms that take into account relativistic effects, and a detailed study of possible degeneracies between environmental effects and post-Newtonian effects~\cite{Speeney:2022ryg}, transient orbital resonances~\cite{PhysRevD.103.124032}, modified gravity, as well as eccentricity~\cite{Yue:2019ozq,Becker:2021ivq} and effects related to the spins of the black holes~\cite{Fairhurst_2020}.

\section*{Acknowledgements}
The authors would like to thank Paolo Pani and Sam Witte for helpful discussions. P.C.\ acknowledges support from the Institute of Physics at the University of Amsterdam. A.C.\ received funding from the Schmidt Futures Foundation. D.G.\ is supported by Spanish MINECO through the Ramon y Cajal programme RYC2020-029184-I starting from 1/09/2022. B.J.K.\ thanks the Spanish Agencia Estatal de Investigaci\'on (AEI, Ministerio de Ciencia, Innovación y Universidades) for the support to the Unidad de Excelencia Mar\'ia de Maeztu Instituto de F\'isica de Cantabria, ref. MDM-2017-0765. T.S. is supported by VILLUM FONDEN (grant no. 37766), the Danish Research Foundation, and the European Union’s H2020 ERC Advanced Grant ``Black holes: gravitational engines of discovery'' grant agreement no. Gravitas–101052587.

\appendix

\section{Modelling dark matter spikes}
\label{app:dd}

We assume that the density profile in \cref{eq:DMspike_profile} extends down to the innermost stable circular orbit (ISCO) of the central IMBH ($r_\mathrm{isco} = 6 G m_1/c^2$, for a central mass $m_1$). We neglect relativistic corrections to the shape of the spike close to $r_\mathrm{isco}$~\cite{Sadeghian:2013laa,Ferrer:2017xwm,Speeney:2022ryg}; at small radii gravitational wave emission dominates over effects from the spike, making the dynamics of the binary largely insensitive to the precise DM density profile close to the merger. We assume that the IMBH has not undergone major mergers with other IMBHs, which would substantially suppress the spike density~\cite{Ullio:2001fb,Merritt:2002vj}. Mergers with lighter compact objects may also affect the precise shape of the density profile, though as argued in Ref.~\cite{Kavanagh:2020cfn} such mergers are unlikely to reduce the DM density by more than a factor of 2. Finally, we assume that the cold DM particle is not self-annihilating; DM annihilation would flatten the central cusp of the DM spike~\cite{Gondolo:1999ef,Bertone:2005hw}.

Guided by $N$-body simulations~\cite{Kavanagh:2020cfn}, we take the maximum impact parameter as $b_\mathrm{max} = \sqrt{q} r$, giving $\Lambda \sim 1/\sqrt{q}$. Dynamical friction traces the DM density in the spike $\rho_\mathrm{CDM}(r, t)$. In the absence of feedback, $\rho_\mathrm{CDM}(r, t)$ would be given by the power-law spike in \cref{eq:DMspike_profile}. However, the injection of energy by the inspiraling compact object leads to a transient depletion of the DM spike. Thus, the DM density profile is time-dependent and can be modelled by evolving the distribution function of particles in the spike during the inspiral, as described in Ref.~\cite{Kavanagh:2020cfn} and implemented in the \texttt{HaloFeedback} code~\cite{HaloFeedback}. The transient depletion of the spike generally reduces the size of the dynamical friction effect, though as demonstrated in Refs.~\cite{Coogan:2021uqv,Cole:2022ucw}, it should still give rise to a dephasing large enough to be observable with future ground- and space-based GW observatories.

For the results in the main text, we generate CDM-dephased waveforms using the \texttt{pydd}~\cite{pydd} code, which uses a broken power law parametrization for the phase evolution, fit to more complete results obtained using \texttt{HaloFeedback}~\cite{HaloFeedback}. For longer signals, such as the 5-year durations studied in \cref{app:duration}, a faster approach to waveform generation is required. In that case, we make use of a surrogate waveform model (also trained on \texttt{HaloFeedback}~\cite{HaloFeedback} results). For further details, see \cref{app:duration}.

We do not include torques for the case of cold collisionless dark matter, because co-rotation and Lindblad torques arise from the fact that the companion is moving coherently with the material in which it is embedded so as to excite orbital resonances. This is not the case for DM spikes, where the particles are not expected to be co-rotating with the binary. Although the halo may be spun up due to the transfer of angular momentum from the companion to the particles, by the time enough energy has been transferred through scatterings for a particle to co-rotate, it will have gained enough energy to become unbound from the halo~\cite{Kavanagh:2020cfn}. We therefore expect the halo to be minimally spun up during the inspiral, and certainly not enough to excite co-rotation or Lindblad torques as is the case for differentially-rotating baryonic disks. 

\section{Modelling Gravitational atoms}
\label{app:ga}

The \textit{ionization power} $\dot{E}_\lab{ion}$ quantifies the energy lost by the binary to the cloud and is defined as \cite{Baumann:2022pkl}
\begin{equation}
    \dot{E}_\lab{ion} = \pm \frac{m_\lab{c}}{\mu}\sum_{\ell', m'} (m'-m) \Omega\, \big|\eta(\epsilon_*)\big|^2\, \Theta(\epsilon_*)\,.
    \label{eqn:ionization-power}
\end{equation}
In this expression, the sum runs over all possible angular momentum states $(\ell',m')$, then $\eta(\epsilon)=\langle\epsilon;\ell'm'|V|n\ell m\rangle$ is the level mixing induced by the gravitational perturbation $V$ of the companion, $\epsilon_*=-\mu\alpha^2/(2n^2)\pm(m'-m)\Omega$, and $\Theta$ is the Heaviside step function. The $\pm$ sign refers to co-/counter-rotating orbits respectively, but $\dot{E}_\lab{ion}$ is guaranteed to be positive in both cases. The backreaction of ionization can be interpreted as dynamical friction \cite{Baumann:2021fkf}.

The \textit{accretion power} is given by
\begin{equation}
    \dot{E}_\lab{acc}=\biggl(\sqrt{\frac{r_2}{Gm_1}}\mp \frac{m}\alpha\biggr)\biggl(\frac{Gm_1}{r_2}\biggr)^{3/2}\frac{\dd m_2}{\dd t}\,,
    \label{eqn:drdt-atom}
\end{equation}
where the accretion rate is equal to
\begin{equation}
    \frac{\dd m_2}{\dd t}=16\pi(Gm_2)^2\rho(\vec r_2)\,,
    \label{eqn:accretion-rate-atom}
\end{equation}
where $\rho(\vec r_2)$ is the local density of the cloud. Finally, the conservation of mass reads
\begin{equation}
    \frac{\dd m_2}{\dd t}+\frac{\dd m_\lab{c}}{\dd t}=-\dot m_\lab{ion},
    \label{eqn:conservation-mass-atom}
\end{equation}
where $\dot m_\lab{ion}$ is the mass lost to ionization, and is given by a formula analogous to (\ref{eqn:ionization-power}), but with the factor $\pm(m'-m)\Omega$ removed. 

We solve numerically the energy balance equation, together with (\ref{eqn:accretion-rate-atom}) and (\ref{eqn:conservation-mass-atom}), for the three quantities $r_2(t)$, $m_2(t)$ and $m_\lab{c}(t)$. The phase of the GW signal is then extracted from $r_2(t)$ as for other environments. The quantities $\dot{E}_\lab{ion}$ and $\dot m_\lab{ion}$ are hard to compute accurately, due to the many overlap integrals of the form $\langle\epsilon;\ell'm'|V|n\ell m\rangle$ that enter the formulae. However, the existence of a universal dependence of those quantities on $\alpha$, $q$ and $m_\lab{c}$ \cite{Baumann:2021fkf} allows us to tabulate their value for a fiducial set of parameters, and simply re-scale them appropriately for every system. The numerical integration of the equations is then very fast.

Notably, this model neglects resonant transitions, which are known to occur at specific orbital frequencies,
\begin{equation}
    (m_a-m_b)\Omega=E_a-E_b,
\end{equation}
where the subscripts $a$ and $b$ refer to the two bound states resonating. Such resonances can lead to the cloud smoothly transitioning from one bound state to another, analogous to the Landau-Zener transition in quantum mechanics \cite{zener1932non,landau1932theorie}. Due to the difference in energy and angular momentum between two bound states, these resonant transitions induce a backreaction on the orbit, which depending on the transition are known as either ``floating'' or ``sinking'' orbits \cite{Baumann:2019ztm}.

\section{Torques from accretion discs}
\label{app:torques}

{In this environment, the behaviour of the drag force term is different with respect to the dark matter case, because of the collisional nature of the medium. The problem of dynamical friction in a gaseous medium was studied in \cite{Ostriker:1998fa} in the case of a straight-line trajectory, and further elaborated in \cite{Kim:2007zb} in the case of a perturber moving on a circular orbit. Both works agree on a vanishing contribution of this effect in the limit of zero relative velocity between the compact object and the medium. Therefore, under the assumption of co-rotation of the companion with respect to the disk, the contribution of dynamical friction as presented in for the dark matter spike can be considered negligible (see also \cite{Barausse:2014pra}).

Instead, the dominant cause of dephasing for compact binaries in accretion disks can be ascribed to gas torques.} 
In the case of dynamical friction, it is the build up of particles, or wake, behind the companion object that slows the orbital velocity and hence drives the inspiral to completion in fewer cycles than in vacuum. 
In the case of accretion disks, it is instead the perturbation of the disk due to the companion object which leads to an asymmetric build up of particles on smaller or larger radii than the orbit of the companion object. These particles then back-react on the black hole and impart gas torques which can either speed up the inspiral (if the build up of particles is on larger radii than the companion) or slow it down (if the build up of particles is on smaller radii). 

Much of the work quantifying this effect has been inherited from planet migration studies \cite{Julian66,Lin79,Goldreich80}, for which there are two classes of migration. Type I migration applies to mass ratios $m_2/m_1<10^{-4}$, where the companion object perturbs the disk linearly. Type II migration applies to mass ratios $m_2/m_1>10^{-4}$, where the companion object perturbs the disk non-linearly, and drives a groove through the disk. This is referred to as \textit{gap-opening}, where gas flows across the gap (usually from larger radii to smaller radii) become an important contribution to the gas torques \cite{Duffell15}.

The analytic expression from \cite{tanaka2002} describes the total net torque on a 2D disk due to a planet orbiting around a star at the centre of the disk:
\begin{equation}\label{eq:2dmig}
    \Gamma_{tot, 2D}=(1.160 + 2.828\alpha)\left(\frac{m_p}{m_c}\frac{r_p\Omega_p}{c}\right)^2\Sigma_pr^4_p\Omega_p^2,
\end{equation}
where $m_p$ is the mass of the planet, $m_c$ the mass of the central star, $r_p$ is the radial separation between the star and the planet, $\Sigma_p$ is the surface density of the (unperturbed) disk at the position of the planet, $\Omega_p$ is the orbital angular velocity of the planet and $\alpha$ is a parameter which describes the radial gradient of the surface density. If $\alpha>0$, then the gradient is negative. The torques are due to the excitation of Lindblad and corotation torques, which arise due to resonances between the pattern speed of the density wave caused by the perturber and the orbital velocity of the gas particles in the disk \cite{Goldreich80}. These torques are present in an entirely Keplerian disc, with the companion also moving at Keplerian velocity, however it is not necessary for this to be the case.

This expression assumes that the perturbations on the disk due to the planet are linear (i.e. Type I migration, small mass ratio between planet and star), that the orbit of the planet is in the plane of the disk and non-eccentric, and also that the disk and perturbations are locally isothermal. It also assumes that the planet is on a fixed orbit and finally, that the disk has no self-gravity and no viscosity.

We use this prescription to describe a pair of black holes with small mass ratio $q=m_2/m_1$ embedded in an accretion disk and write the total net torque \textit{on the secondary black hole}, hence the opposite sign to \cref{eq:2dmig} which describes the torques on the gas which then back-react onto the black hole. Neglecting the pre-factor from \cref{eq:2dmig}, the net torque on the secondary black hole is \cite{Derdzinski_2020}
\begin{equation}\label{eq:torque2}
    T_0=-\Sigma(r)r^4\Omega^2q^2 M^2,
\end{equation}
which is the expression we use in \cref{eq:torque}.

Although this analytical description makes many approximations and was calculated in the context of planet migration, 2D hydrodynamical numerical simulations results have shown that it captures the effects of the gas torques on a secondary black hole being driven towards a central black hole due to gravitational wave emission to within a factor of a few, although the level of agreement is very sensitive to the parameters of the binary and the disk \cite{Derdzinski:2018qzv,Derdzinski_2020}. We choose to use the analytic expression as a first approximation to the gas torques, sufficient to distinguish between different environments. It will also facilitate fast production of gravitational waveform templates in the presence of an accretion disk, which will be necessary for efficient parameter estimation in \cref{sec:PE}. 

However, in order to learn about the properties of accretion disks from gravitational waves, a more complete description of the interplay between the secondary black hole and the accretion disk will be vital, and we expect this to be informed by improvements to numerical simulations.

One interesting subtlety found in \cite{Derdzinski_2020} was that resolving the gas torques inside the Hill sphere region surrounding the secondary black hole displays a clear asymmetry of gas in front and behind the companion. For mass ratios $q>3\times10^{-4}$, this can change the overall sign of the torques. For $q<3\times10^{-4}$, this effect is shown to be small and the analytic prescription does better, modulo sensitivity to disk parameters which have only been simulated within given ranges. We will only report negative torques on the black hole, however, if the sign of the torques is in fact positive due to a combination of disk and binary factors, we would expect distinction of accretion disks versus other environments to be even more tractable.

We do not include the effects of accretion onto the companion object, as these were shown to be small in the gas-only simulations conducted in \cite{Derdzinski_2020}. However note that this may not hold for all regions of the parameter space, and may not hold when radiation is included in the simulations. Furthermore, the impact of turbulence \cite{Zwick2021}, eccentricity, relaxing the locally isothermal disk model, as well as 3D effects could all impact the amount of dephasing that can be expected from these systems. For now, we provide a proof of concept for distinguishing accretion disks from dark matter environments with this simple analytical model, and aim to update the modelling of the accretion disk as numerical simulations progress.

\section{Dephasing}
\label{app:dephasing}

Each of the energy loss terms can be written instead in terms of the rate of change of the separation of the binary
\begin{equation}
    \dot{r} =  \frac{2r^2\dot{E}}{ G m_1 m_2}=\dot{r}_{\rm GW} + \dot{r}_{\rm env}.
\end{equation}
The phase of the signal is then related to this by:
\begin{equation}
\label{eq:fKep}
    f(t)=\frac{1}{\pi}\sqrt{\frac{G(m_1 + m_2)}{r(t)^3}}\,, \quad \Phi(f) = \int_f^{f_\mathrm{ISCO}} \frac{\mathrm{d}t}{\mathrm{d}f^\prime} f^\prime \,\mathrm{d}f^\prime\,.
\end{equation}
The observable effect on the waveform, i.e.\ the difference in the number of cycles $N_{\rm cyc}$ from a given reference frequency until ISCO (or merger) between a given environment and the vacuum case, is then the \textit{dephasing}:
\begin{equation} \label{eq:dephasing_definition}
    \delta\Phi = 2\pi (N_{\rm cyc, V} - N_{\rm cyc, env}) = \Phi_{\rm V} - \Phi_{\rm env}\,.
\end{equation}

This phase shift with respect to an inspiral in vacuum enters into the gravitational waveform via its second derivative with respect to time, and here we use the Newtonian order expression
\begin{align}
    h_0(f) = \frac{1}{2} \frac{4 \pi^{2/3} G^{5/3} \mathcal{M}^{5/3} f^{2/3}}{c^4} \sqrt{\frac{2\pi}{\ddot{\Phi}}}\,,\qquad \text{where} \quad
    \ddot{\Phi} = 4\pi^2 f \qty(\dv{\Phi}{f})^{-1}.
    \label{eq:strain_and_phase}
\end{align}

\begin{figure}[t]
     \centering
         \includegraphics[width=0.8\textwidth]{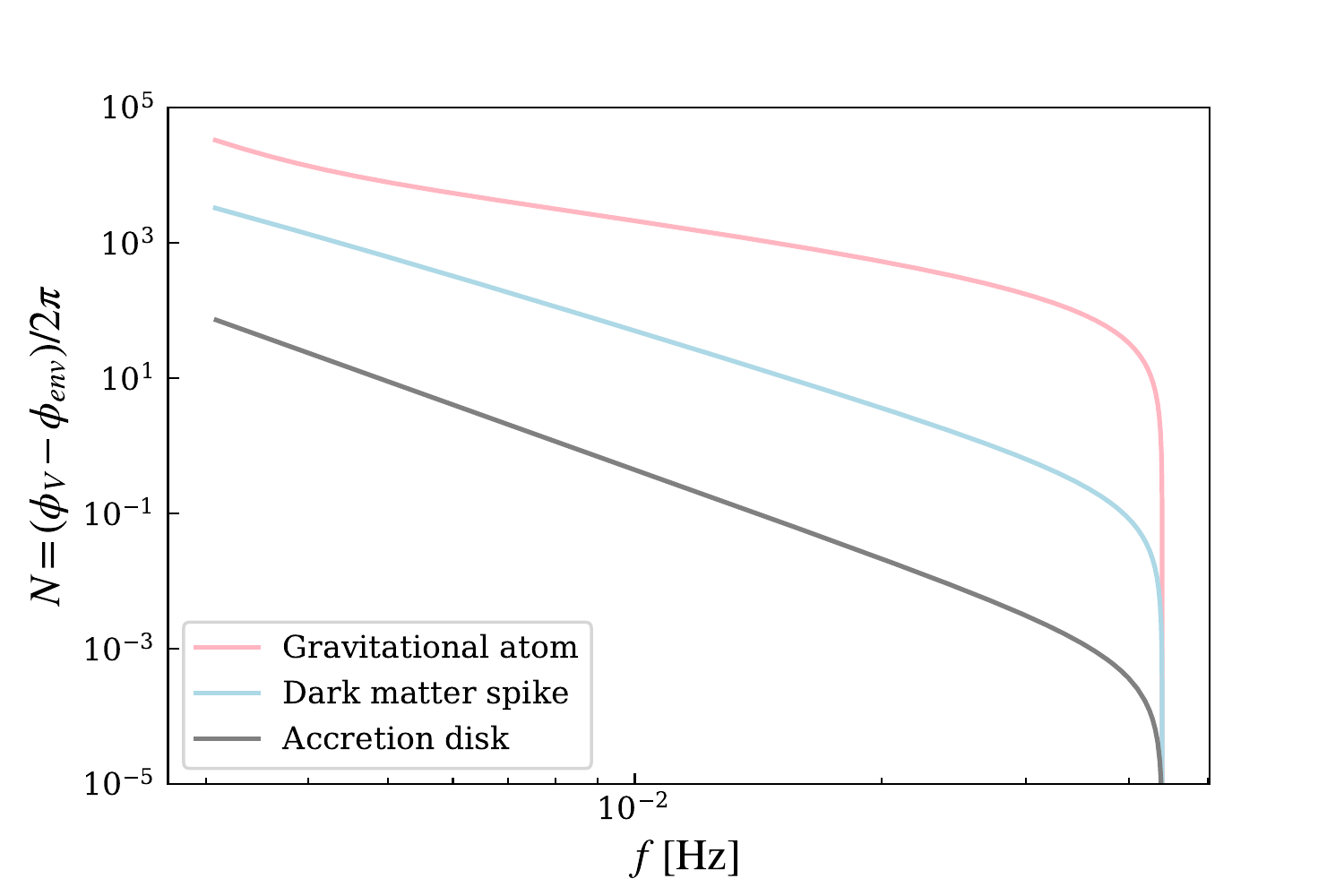}

        \caption{Dephasing with respect to naive vacuum signals, using the models for the frequency evolution in \cref{sec:elosses} which we use for parameter estimation in \cref{sec:PE}.}
        \label{fig:dephasing}
\end{figure}

\section{Match integral, SNR, and faithfulness}
\label{sec:match}

The likelihood maximized over extrinsic parameters between the signal $d(t)$ and the template $h_\theta(t)$ with intrinsic parameters $\theta$ is
\begin{equation}\label{eq:maxl}
    p(d|h) = \exp\left(\frac{\left< h_\theta | d \right>_\mathrm{max}^2}{2\left< h_\theta | h_\theta \right>}\right) \, ,
\end{equation}
where the inner product between the signal and template in the numerator is given by
\begin{equation}
    \left<h|d\right>_\mathrm{max} = \max_{t_c} \operatorname{FT}\left( 4 \left| \frac{\tilde{h}_\theta^*(f) \tilde{d}(f)}{S_n(f)} \right| \right)(t_c) \, .
\end{equation}
Here $t_c$ is the time at coalescence and $\operatorname{FT}(g)(\cdot)$ is the Fourier transform of the function $g(\cdot)$. We assume that our time series data is a linear combination of the signal plus the detector noise, which is Gaussian and given by the power spectral density noise curve as made publicly available in \cite{Robson_2019} by the LISA consortium. We maximise over extrinsic parameters of the binary, namely the phase of coalescence $\phi_c$ (by taking the absolute value), the time of coalescence $t_c$ (by taking the fast Fourier transform), and the luminosity distance $d_L$ to the binary (with the normalisation factor in \cref{eq:maxl}), all according to \cite{Owen_1996}.
We thus assess our ability to reconstruct the intrinsic parameters of the binary defined in the detector frame, which are the chirp mass $\mathcal{M}_c$, mass ratio $q$, and the parameters that describe each of the environments $\theta_\mathrm{env}$. Prior ranges for these parameter values are given in \cref{tab:priors}. We use noise-free (i.e., Asimov~\cite{Cowan:2010js}) signals that span 1 year's worth before the time of coalescence, which we take to be at the time corresponding to the ISCO frequency, and we place the mergers at a distance such that the accumulated signal-to-noise ratio (SNR) over that time is 15, where the optimal SNR of a signal $d$ is
\begin{equation}
    \mathrm{SNR} = \sqrt{\left<d|d\right>}.
\end{equation}

\begin{figure}[h!]
    \centering
    \includegraphics[width=0.8\textwidth]{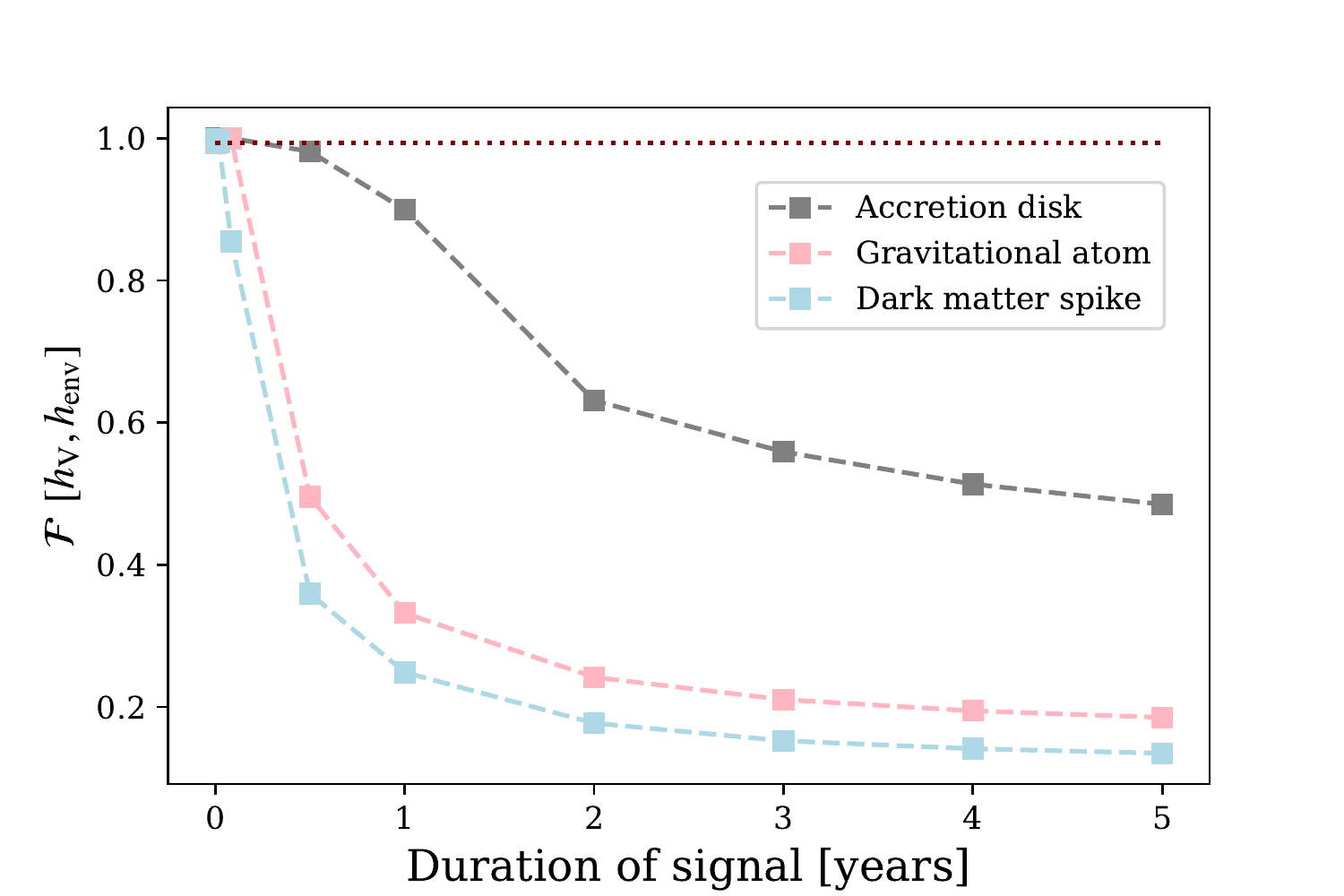}
    \caption{Faithfulness of environmental signal waveform compared to the best-fit vacuum system as a function of the duration of the signal, measured in years before merger, for our benchmark accretion disk (grey), dark dress (light blue) and gravitational atom (light pink).}
    \label{fig:faith}
\end{figure}

The \textit{faithfulness} of a given system's waveform with respect to another is given by \cite{Maselli_2022}
\begin{equation}\label{eq:faith}
    \mathcal{F}[h_\theta,d]=\frac{\left<h_\theta|d\right>}{\sqrt{\left<h_\theta|h_\theta\right>\left<d|d\right>}}.
\end{equation}
In \cref{fig:faith} we show the faithfulness of the signal waveform with respect to its best-fit vacuum system, as a function of the duration of the signal. All durations are measured backwards from ISCO frequency and the distance to the source is fixed such that the SNR of the 1 year signal is 15. The faithfulness degrades substantially for the gravitational atom and the dark dress for durations longer than 1 month, whilst the accretion disk faithfulness degrades more gradually as expected. The threshold on the faithfulness required to distinguish between environments is $\mathcal{F}=0.99$, shown by the dotted dark red horizontal line. This is calculated for an SNR of 15 as laid out in \cite{Chatziioannou_2017}, and we choose a dimensionality of 4 in the parameters which makes the threshold conservative for the accretion disk system which is only described by 2.

\section{Fitting dark dresses and gravitational atoms  with accretion disk templates}

\begin{figure}[t]
    \centering
    \includegraphics[width=0.45\textwidth]{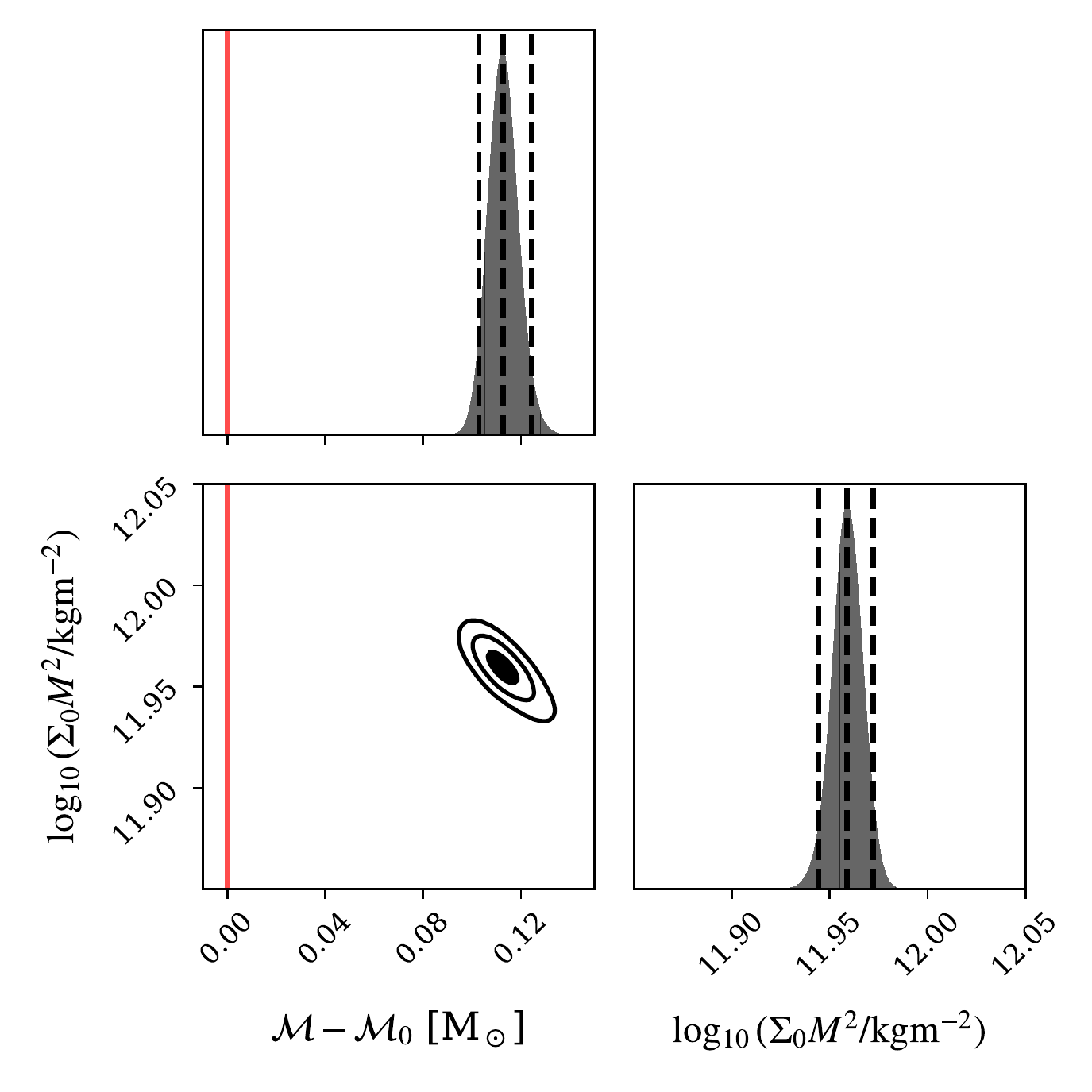}
    \includegraphics[width=0.45\textwidth]{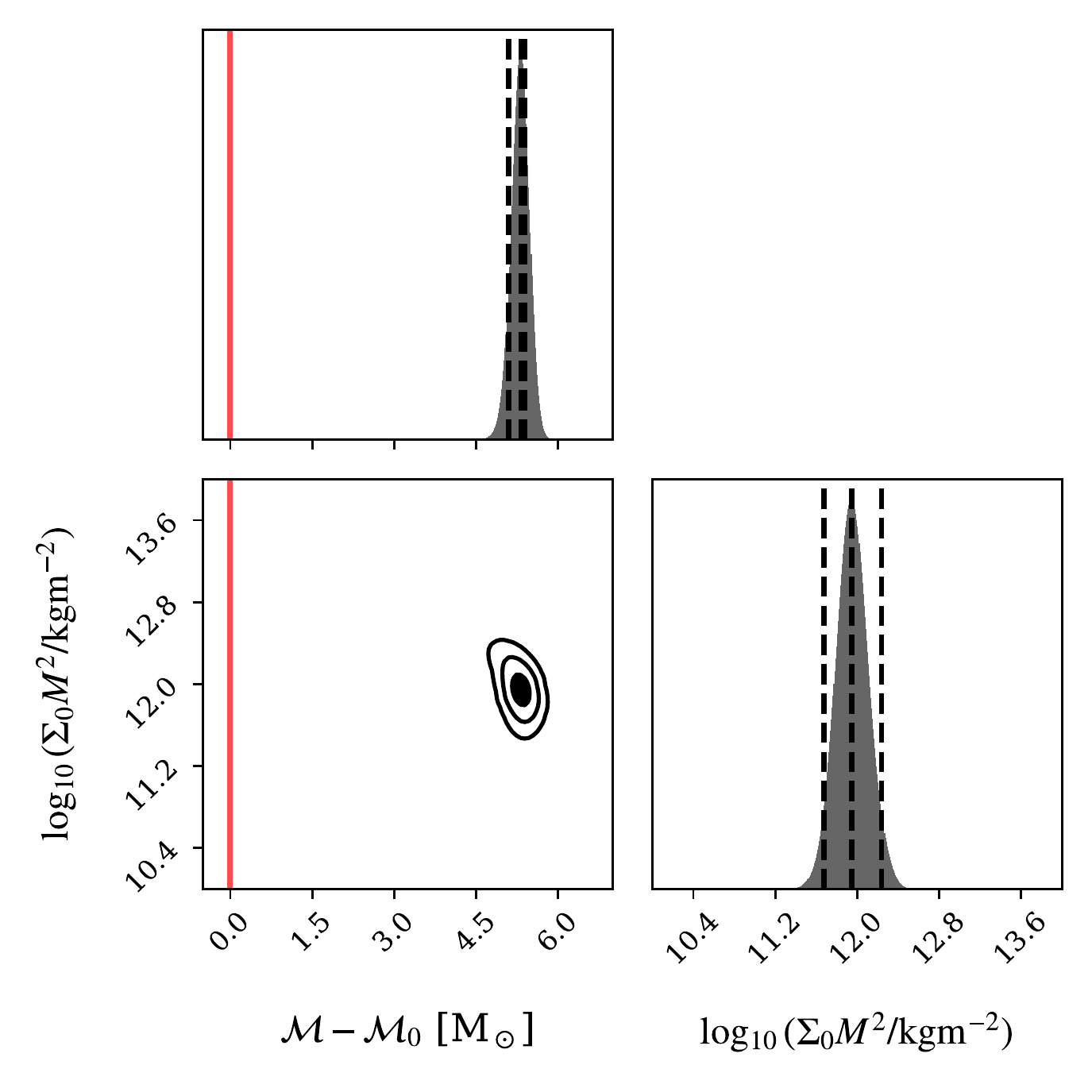}
    \caption{Posterior distribution for the chirp mass and environmental parameters when fitting a dark dress signal (left), and a gravitational atom signal (right) with an accretion disk template. The true chirp mass, $\mathcal{M}_0$, is indicated by the red vertical line at $\mathcal{M}-\mathcal{M}_0=0$.}
    \label{fig:ad-fits-2}
\end{figure}

The pair of systems with the lowest Bayes factors are those of fitting an accretion disk signal with a dark dress or gravitational wave template. Note that all parameter estimation runs including accretion disks involve fixing the mass ratio because it is degenerate with the surface density and Mach number combination in our set-up. This drives the better reconstruction of the remaining parameters, although the accretion disk template is still severely disfavoured with respect to the correct model. However, it is still interesting to observe that the wrong template can fit the signals well in these cases. As an example, the posteriors for fitting our benchmark dark dress and gravitational atom signals with an accretion disk template are shown in \cref{fig:ad-fits-2}. In both cases, the parameter reconstruction infers a biased-high chirp mass as well as a large value for $\Sigma_0 M^2$, mimicking the speed-up of the inspiral due to dynamical friction and ionization/accretion respectively.

\section{Parameter estimation with different waveform durations including the use of surrogate models}\label{app:duration}

With 1 year's worth of data, as shown in \cref{fig:correct}, the parameters of the dark dress system are only partially measured, whilst the parameters of the gravitational atom are measured to incredible precision. We show here that increasing the duration of the signal to 5 years for the dark dress (which would be the best-case scenario for LISA observations) leads to well-converged posteriors for all parameters. 
\begin{figure}[t!]
    \centering
    \includegraphics[width=0.8\textwidth]{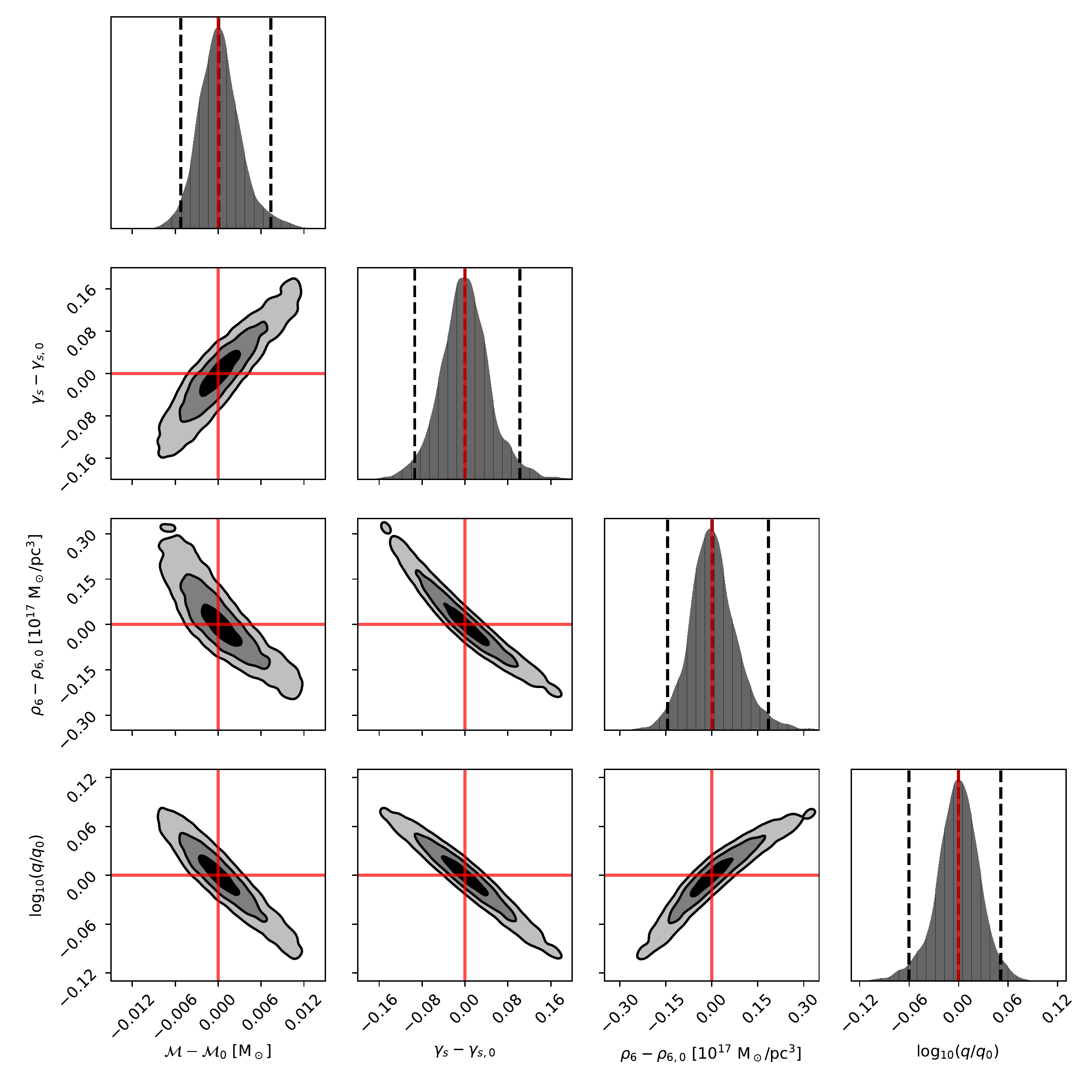} 
    \caption{
    Posterior distributions for the chirp mass, mass ratio and environmental parameters of a dark dress with an SNR of 15 and using 5 year's worth of data. 
    }
    \label{fig:dur}
\end{figure}
\begin{figure}[h]
    \centering
    \includegraphics[width=0.8\textwidth]{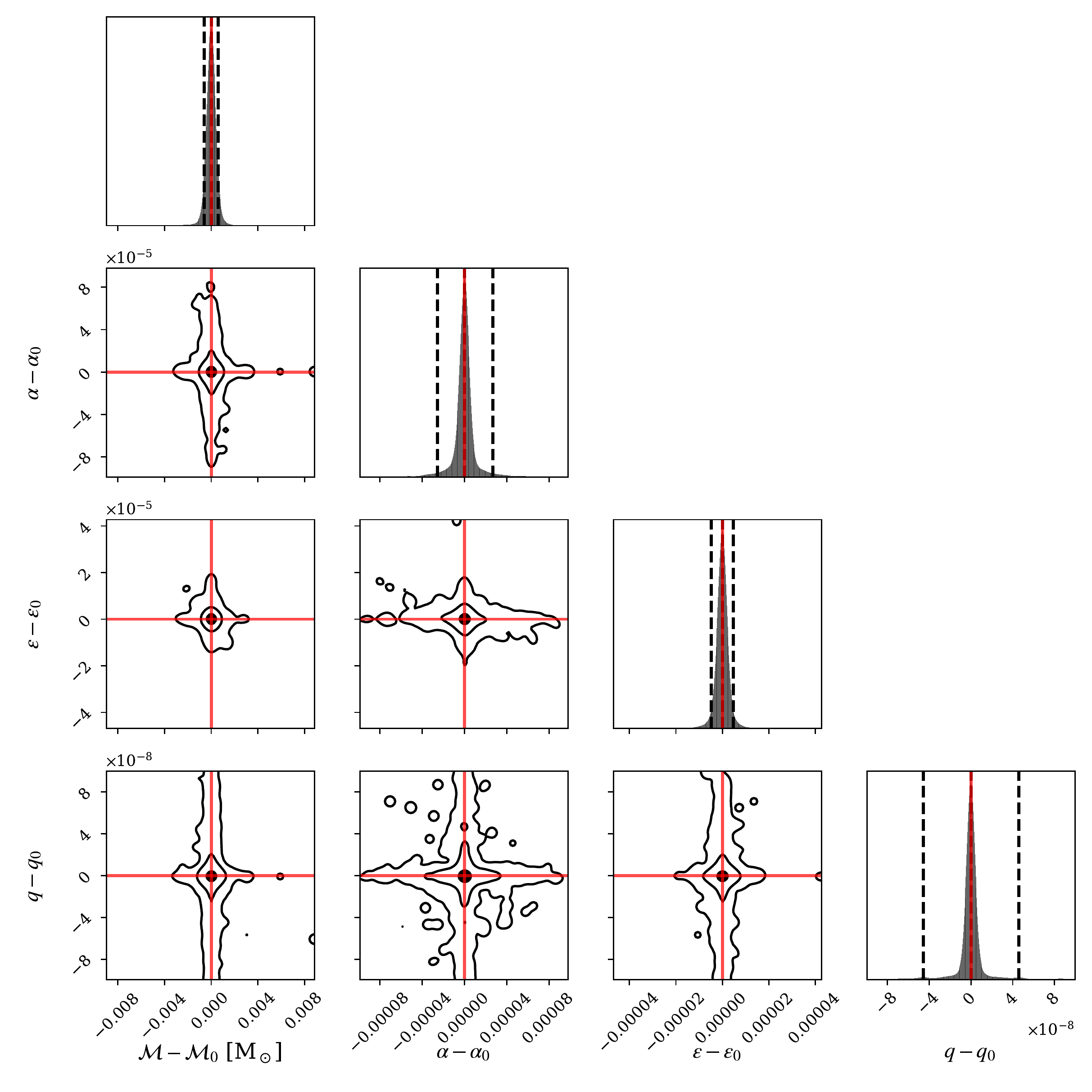}
    \caption{Posterior distributions for the chirp mass, mass ratio and environmental parameters of a gravitational atom with an SNR of 15 and using 1 month's worth of data. The true values are indicated by the red lines.}
    \label{fig:dur2}
\end{figure}
For long waveforms, the analytical approximation introduced in \cite{Coogan:2021uqv} becomes inadequate to capture the evolution of the phase $\Phi(f)$. Instead of devising a new analytical approximation for long waveforms, we introduce here a {\it surrogate model} for the phase evolution, trained on the fiducial waveforms produced by the \texttt{HaloFeedback} code.

We generate a dataset split between 852 train and 100 test `points' that represent the evolution of dressed binaries initialised 25 years before the coalescence of a vacuum system. Their intrinsic parameters correspond to our model's input features $\bm{X} = \{\log_{10}(m_1),\, \log_{10}(m_2),\, \log_{10}(\rho_6),\, \gamma_s\}$, and were sampled to cover the intended application range 
$\log_{10}(m_1/\mathrm{M}_\odot) = \mathcal{U}(2.9 , 6)$, $m_2/\mathrm{M}_\odot = \mathcal{U}(1 , 15)$, $\log_{10}(\rho_6/(\mathrm{M_\odot/pc^3})) = \mathcal{U}(13,19)$, and $\gamma_s = \mathcal{U}(2,2.55)$. After ignoring approximately the first 15 years of the simulation to avoid initial transients (see a discussion in \cite{Kavanagh:2020cfn}), we extract the phase evolution $\Phi(f)$ until the merger frequency $f_c$, taken at the binary's ISCO radius. From that we construct the dephasing according to \cref{eq:dephasing_definition}.

For any choice on the input parameters, the logarithm of the dephasing curve
achieves a similar functional form after
the linear frequency transformation
\begin{equation}
  x(f; m_1, m_2) = \frac{f-f_{c}}{f_{10yr}(m_1, m_2)} -f_{c} \in [0, 1] \, ,
\end{equation}
where $f_{10yr}(m_1, m_2)$ is the GW frequency emitted by a vacuum binary 10 years before merger.
We transform the dephasing curves $\log_{10}{\Delta \Phi_{\to c}(x)}$ using dimensionality reduction, with a greedy basis-construction algorithm \cite{RomPy,Field_2011}, and ensure that the maximum reconstruction errors 
falls below a threshold of $10^{-6}$, resulting in a reduced, 24-dimensional basis.

To learn the mapping between the intrinsic parameters $\bm{X}$ and the low-dimensional representation of the dephasing
$\bm{Y}$, we implement Gaussian process regression (GPR) \cite{gpbook,murphy2013machine}, with a linear combination of a radial basis kernel and a white Kernel, such that for two input parameters $\bm{X}, \bm{X}'$,
\begin{equation}
  k(\bm{X}, \bm{X}') = c_0 e^{-\frac{\lVert \bm{X} - \bm{X}' \rVert^2}{2\lambda^2}} + \sigma^2 \delta_{\bm{X}, \bm{X}'} \, ,
\end{equation}
where $\delta_{\cdot, \cdot}$ is the Kronecker delta function. The hyperparameters $(c_0, \lambda, \sigma^2)$ are tuned over many iterations during training using scikit-learn \cite{scikit-learn} to maximize the log-likelihood of the GP. After the GPR is conditioned, we can evaluate it at any point of the parameter space within the training range to find basis function coefficients, and through them reconstruct the dephasing curve $\log_{10}{\Delta \Phi_{\to c}(x)}$, and finally the phase $\Phi(f)$. In order to validate the phase evolution obtained with the surrogate model, we have compared it with a test set of waveforms generated with the \texttt{HaloFeedback} code. We found that the surrogate model improves by one order of magnitude the accuracy in the calculation of the phase evolution in frequency with respect to the analytical approximation in 
\cite{Coogan:2021uqv}. In particular, it more effectively captures the evolution of the phase around the `break' region where the analytical approximation interpolates between two power laws.

In \cref{fig:dur}, we show the 4D posteriors obtained with the surrogate model, in 1D and 2D. All posteriors are all well converged and contained within the priors. We note that with these longer waveforms it becomes possible to accurately measure $\gamma_s$. This is a consequence of the fact that with 5 years worth of data, the break frequency of the dephasing is observed, which is a distinctive phase of the inspiral evolution that carries key information on the properties of the dark matter spike. 

We also show in \cref{fig:dur2} that with just one month of data the posterior distributions for a gravitational atom signal are still measured to excellent precision. The $95\%$ credible intervals for the parameters are $\mathcal{M}_c = 398.099^{+0.0006}_{-0.0006}$, $q = 0.0001 \pm{10^{-8}}$, $\alpha = 0.2 \pm{10^{-5}}$ and $m_c/m_1 = 0.01\pm{10^{-6}}$ (which we denote as $\epsilon$ in \cref{fig:dur2}). We keep the SNR fixed at 15 for all durations, meaning that the source is further away for the dark dress and closer for the gravitational atom in Figs. \ref{fig:dur} and \ref{fig:dur2}.

\bibliography{main.bib}

\end{document}